\documentclass[twocolumn,english,amsart,showpacs,preprintnumbers,amsmath,amssymb,floatfix]{revtex4-1}
\usepackage{tikz,xcolor}
\usepackage[colorlinks = true,
linkcolor = blue,
urlcolor  = blue,
citecolor = blue,
anchorcolor = blue]{hyperref}

\definecolor{lime}{HTML}{A6CE39}
\DeclareRobustCommand{\orcidicon}{%
	\begin{tikzpicture}
		\draw[lime, fill=lime] (0,0)
		circle [radius=0.16]
		node[white] {{\fontfamily{qag}\selectfont \tiny ID}};
		\draw[white, fill=white] (-0.0625,0.095)
		circle [radius=0.007];
	\end{tikzpicture}
	\hspace{-2mm}
}

\foreach \x in {A, ..., Z}{%
	\expandafter\xdef\csname orcid\x\endcsname{\noexpand\href{https://orcid.org/\csname orcidauthor\x\endcsname}{\noexpand\orcidicon}}
}

\usepackage[T1]{fontenc}
\usepackage[latin9]{inputenc}
\usepackage{color}
\usepackage{array}
\usepackage{amstext}
\usepackage{graphicx}
\usepackage{esint}
\usepackage{rotating}
\usepackage{appendix}
\usepackage{float}

\usepackage[font=small,labelfont=bf]{caption}
\usepackage{xcolor}
\usepackage{ulem}

\makeatletter

\@ifundefined{textcolor}{}
{%
	\definecolor{BLACK}{gray}{0}
	\definecolor{WHITE}{gray}{1}
	\definecolor{RED}{rgb}{1,0,0}
	\definecolor{GREEN}{rgb}{0,1,0}
	\definecolor{BLUE}{rgb}{0,0,1}
	\definecolor{CYAN}{cmyk}{1,0,0,0}
	\definecolor{MAGENTA}{cmyk}{0,1,0,0}
	\definecolor{YELLOW}{cmyk}{0,0,1,0}
}

\@ifundefined{definecolor}
{\usepackage{color}}{}
\@ifundefined{definecolor} 
{\usepackage{color}}{}
\makeatother
\usepackage{babel}

\allowdisplaybreaks

\begin{document}
	

	\title{Nucleon Form Factors from GPDs}

	\author{Leila Ghasemzadeh$^{1}$\orcidA{}}
	\email{leila.ghasemzadeh71@gmail.com}
	\author{Pegah Sartipi YarAahmadi $^{2}$\orcidC{}}
	\email{Pegah.sartipi.yar@gmail.com}
	\author {Fatemeh Arbabifar$^{3}$\orcidF{}}
    \email{F.Arbabifar@cfu.ac.ir}
	\author{Nader Morshedian$^{4}$\orcidE{}}
    \email{nmorshed@aeoi.org.ir}
	\author{Shahin Atashbar Tehrani$^{5,6}$\orcidB{}}
	\email{atashbar@ipm.ir}
	 
\affiliation {
$^{(1)}$Eram High Education Institute of Shiraz, Postal code 7195746733, Shiraz, Iran\\
$^{(2)}$Department of Physics, Ayatollah Boroujerdi University, P.O.Box 69737-69199, Boroujerd, Iran\\
$^{(3)}$Department of Physics Education, Farhangian University, P.O.Box 14665-889, Tehran, Iran.   \\
$^{(4)}$Plasma and Nuclear Fusion Research School, Nuclear Science and Technology Research Institute, P.O.Box 14399-51113, Tehran, Iran.\\
$^{(5)}$School of Particles and Accelerators, Institute for Research in Fundamental Sciences (IPM), P.O.Box 19395-5531, Tehran, Iran.\\
$^{(6)}$Department of Physics, Faculty of Nano and Bio Science and Technology, Persian Gulf University, 75169 Bushehr, Iran.  }

	\date{\today}

	%
	%

	%
	\begin{abstract}\label{abstract}
In this study, we introduce a novel ansatz for Generalized Parton Distributions (GPDs), named GSAMA24. This ansatz aims to provide a more accurate and comprehensive description of the internal structure of hadrons by incorporating advanced parameterizations and fitting techniques. We compare the performance of the GSAMA24 ansatz with three established models: the Extended Regge (ER), Modified Gaussian (MG), and M-HS22 ansatz. The GSAMA24 ansatz is designed to address limitations observed in previous models by offering improved flexibility in the ( $t$ )-dependence and skewness parameter $\xi$. Our analysis involves fitting the GSAMA24 ansatz to experimental form factor data and evaluating its predictive power against the ER, MG, and M-HS22 models. The comparison is based on key metrics such as the accuracy of form factor predictions, the consistency with known GPD properties, and the computational efficiency of the fitting process. Results indicate that the GSAMA24 ansatz provides a superior fit to the experimental data, particularly in the high momentum transfer region, where it outperforms the ER and MG models. Additionally, the GSAMA24 ansatz demonstrates better agreement with the theoretical expectations of GPD behavior compared to the M-HS22 model. These findings suggest that the GSAMA24 ansatz is a promising tool for future studies of hadronic structure and could significantly enhance our understanding of the spatial and momentum distributions of quarks and gluons within hadrons.

	\end{abstract}
	%

	
	\maketitle
	\tableofcontents{}

	%

	\section{INTRODUCTON}\label{sec:sec1}
	Over the last ten years, extensive research into Generalized Parton Distributions (GPDs) has demonstrated their effectiveness in characterizing the composition of hadrons. GPDs extend our knowledge of parton distribution functions and offer a comprehensive repository of data on the hadrons' composition. The possibility of experimentally measuring GPDs is enabled by the collinear factorization theorems applicable to hard exclusive reactions. Generalized parton distributions (GPDs) are considered one of the essential techniques for studying the structure of nucleons~\cite{Muller:1994ses,Ji:1996ek,Radyushkin:1997ki,Ji:1996nm,Collins:1996fb,Diehl:2003ny}.
	
	One of the key applications of GPDs is in the study of nucleon form factors. Nucleon form factors are quantities that describe the distribution of charge and magnetization within a nucleon. By connecting GPDs to nucleon form factors, we can gain a deeper understanding of the underlying structure of nucleons and the dynamics of the strong force.
	
	Recent studies have focused more on the structure of nucleons through the lens of Parton Distribution Functions (PDFs) and GPDs. These GPDs are crucial for revealing the distribution of quarks within hadrons and are closely linked to the Dirac and Pauli form factors, providing valuable information about the electromagnetic properties of nucleons \cite{Perdrisat:2006hj,Arnold:1980zj,JeffersonLabHallA:2011yyi,Gramolin:2021gln,Punjabi:2005wq,Rosenbluth:1950yq,Gayou:2001qt,Halzen}. The form factors for valence quarks are symbolized by 
	 $H_{q}(x,\xi, t)$ and $E_{q}(x,\xi,t)$  \cite{Guidal:2004nd,Nikkhoo:2015jzi,RezaShojaei:2016oox}, where the variable,$t$ denotes the momentum transfer,
	$x$ represents the average longitudinal momentum fraction carried by a parton, and the skewness parameter 
	,$\xi$ quantifies the asymmetry in the parton's longitudinal momentum distribution \cite{Radyushkin:2011dh,Guidal:2013rya,Bhattacharya:2019cme,HajiHosseiniMojeni:2022okc}.\\
In this paper, we present our methodology of identifying an appropriate ansatz and integrating it with a corresponding PDF. We compare form factors using the MG~\cite{Selyugin:2009ic}, ER~\cite{Guidal:2004nd}, and M-HS22~\cite{Vaziri:2023xee} models.

Our proposed ansatz, GSAMA24, offers a more accurate description of hadrons' internal structure through advanced parameterizations. It provides improved flexibility in ( t )-dependence and the skewness parameter $ \xi $, allowing for a precise fit to experimental data, especially in the high momentum transfer region. Compared to the MG~\cite{Selyugin:2009ic}, ER~\cite{Guidal:2004nd}, and M-HS22~\cite{Vaziri:2023xee} models, GSAMA24 shows better agreement with theoretical expectations and enhances computational efficiency. Our analysis confirms GSAMA24's superior fit and predictive power.

	This paper is organized as follows: Section \ref{sec:sec2} provides a concise overview of generalized parton distributions and their characteristics. Section \ref{sec3} delves into the ramifications of the chosen ansatz. In In this paper, we present our methodology of identifying an appropriate ansatz and integrating it with a corresponding PDF. We compare form factors using the MG~\cite{Selyugin:2009ic}, ER~\cite{Guidal:2004nd}, and M-HS22~\cite{Vaziri:2023xee} models.

\section{Generalized Parton Distributions and Nucleon Form Factors}\label{sec:sec2}
GPDs provide a unified description of the nucleon structure in terms of parton distributions and form factors. By studying GPDs, we can gain insights into the spatial distribution of quarks and gluons inside the nucleon.

To calculate nucleon form factors using GPDs, one typically starts by parametrizing the GPDs based on experimental data and theoretical models. These parametrizations can then be used to calculate the form factors by performing integrals over the GPDs.

It's important to note that the calculation of nucleon form factors using GPDs can be a complex and computationally intensive process. However, with advancements in theoretical techniques and computational tools, researchers are making significant progress in this area.

To acquire a three-dimensional portrayal of the nucleon's spatial distribution of partons in the transverse plane, researchers conduct deep virtual Compton scattering experiments, subsequently applying a Fourier transform to the 
t-dependence of GPDs \cite{Burkardt:2002hr, Burkardt:2000za, Ralston:2001xs, Belitsky:2003nz, SattaryNikkhoo:2018odd}.

The nucleon Dirac and Pauli form factors 
\( F_{1}(t) \) and \( F_{2}(t) \) are given by

\begin{equation}
F_{i}(t)=\sum_{q}e_{q}F_{i}^{q}(t).
\end{equation}
Using the valence quark GPDs \( H \) and \( E \) through the following sum rules for their flavor components:
\begin{equation}\label{eq:1}
F_{1}^{q}(t)=\int_{-1}^{1}dx H^{q}(x,\xi,t).
\end{equation}
\begin{equation}\label{eq:2}
F_{2}^{q}(t)=\int_{-1}^{1}dx E^{q}(x,\xi,t).
\end{equation}
Here, \( q \) refers to two quarks, \( u \) and \( d \).

When the momentum is transverse and falls within the space-like region, the skewness parameter \( \xi \) is zero. Furthermore, the integration region can be reduced to the \( 0 < x < 1 \) interval, introducing the nonforward parton densities \cite{Radyushkin:1998}.
\begin{equation}
\mathcal{H}^{q}(x,t)=H^{q}(x,0,t)+H^{\bar{q}}(-x,0,t).
\end{equation}
\begin{equation}
\mathcal{E}^{q}(x,t)=E^{q}(x,0,t)+E^{\bar{q}}(-x,0,t).
\end{equation}
Given the conditions mentioned, we can write Eq.\ref{eq:1} and Eq.\ref{eq:2} in the following manner \cite{Guidal:2004nd}:
\begin{equation}\label{eq:6}
F_{1}^{q}(t) = \int_{0}^{1}dx\mathcal{H}^{q}(x,t).
\end{equation}
\begin{equation}\label{eq:7}
F_{2}^{q}(t) = \int_{0}^{1}dx\mathcal{E}^{q}(x,t).
\end{equation}
When \( t \rightarrow 0 \), the \( \mathcal{H}^{q} \) functions satisfy the following relations:
\begin{equation*}
\mathcal{H}^{u}(x,t=0)=u_{v}(x).
\end{equation*}
\begin{equation*}
\mathcal{H}^{d}(x,t=0)=d_{v}(x). 
\end{equation*}
While connecting them with the standard valence quark densities in the nucleon.

The \( t = 0 \) limit of the  \( \mathcal{E}^{q}(x, t) \) distributions exists, but the magnetic densities \( \mathcal{E}^{q}(x,0) \equiv \mathcal{E}^{q}(x) \) cannot be directly expressed in terms of any known parton distribution; they contain new information about the nucleon structure. Additionally, when \( x \rightarrow 1 \) function should include more powers of \( (x-1) \), the \( \mathcal{E}(x) \) than the \( \mathcal{H}(x) \) function to produce a quicker reduction with \( t \) \cite{Guidal:2004nd,Selyugin:2009ic}. Therefore, this function can be defined for \( u \) and \( d \) quarks as follows by introducing the normalization integral below \cite{Radyushkin:1998}:

\begin{equation}
\kappa_{q}=\int_{0}^{1}dx\mathcal{E}_{q}(x).
\end{equation}
The normalization integrals are constrained by the requirement that the values \( F^{p}_{2}(t = 0) \) and \( F^{n}_{2}(t = 0) \) are equal to the anomalous magnetic moments of the proton and neutron. This gives:
\begin{equation}\label{eq:9}
\kappa_{u}=2\kappa_{p}+\kappa_{n}\approx+1.673,
\end{equation}
\begin{equation}\label{eq:10}
\kappa_{d}=\kappa_{p}+2\kappa_{n}\approx-2.033.
\end{equation}
While \( \kappa_{n}=F_{2}^{n}(0)=-1.913 \) and \( \kappa_{p}=F_{2}^{p}(0)=1.793 \). It is important to note that the normalization integrate \( \int_{0}^{1}dx\mathcal{H}_{q}(x,0) \) is equal to \( F_{1}^{p}=1 \) for the proton and \( F_{1}^{n}=0 \) for the neutron. Given these definitions and constraints, we can define \( \mathcal{E}_{q}(x) \) functions as follows:
\begin{equation*}
\mathcal{E}_{u}(x)=\frac{\kappa_{u}}{N_{u}}(1-x)^{\eta_{u}}u_{v}(x).
\end{equation*}
\begin{equation}\label{eq:11}
\mathcal{E}_{d}(x)=\frac{\kappa_{d}}{N_{d}}(1-x)^{\eta_{d}}d_{v}(x).
\end{equation}
Which \( N_{q} \)s are the normalization factors and can be determined for the \( u \) quark and the \( d \) quark, respectively as follows \cite{Guidal:2004nd}:
\begin{equation}
N_{u}=\int_{0}^{1}dx(1-x)^{\eta_{u}}u_{v}(x).
\end{equation}
\begin{equation}
N_{d}=\int_{0}^{1}dx(1-x)^{\eta_{d}}d_{v}(x).
\end{equation}
Up to this point in the paper, we have explained how to calculate the form factors of nucleons using GPDs. Next, we will introduce some of the reputable ansatzes that have been proposed in the past by various groups. Following that, we will introduce the GSAMA24 ansatz.

	\begin{figure*}
	\includegraphics[clip,width=0.45\textwidth]{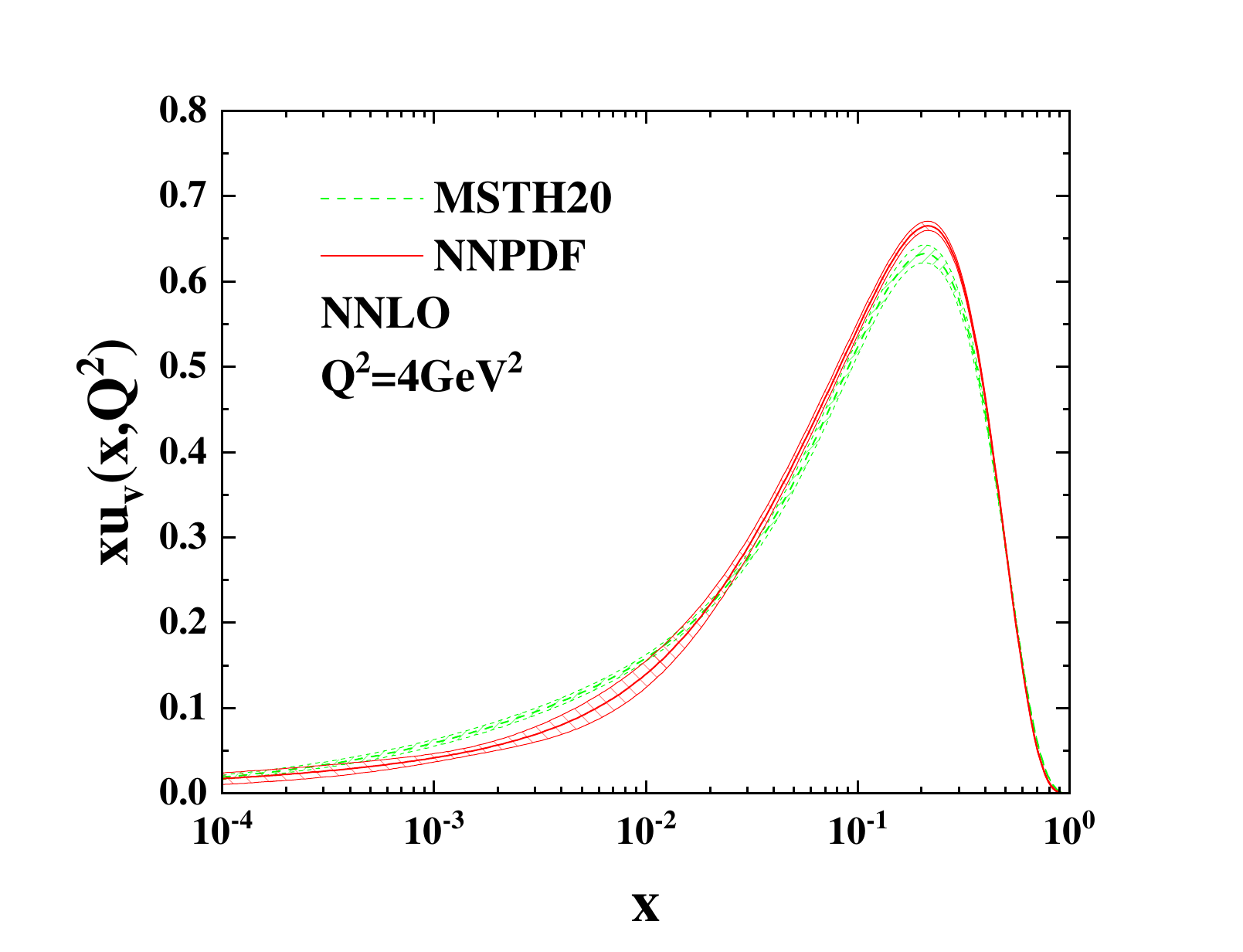 }
	\hspace*{2mm}
	\includegraphics[clip,width=0.45\textwidth]{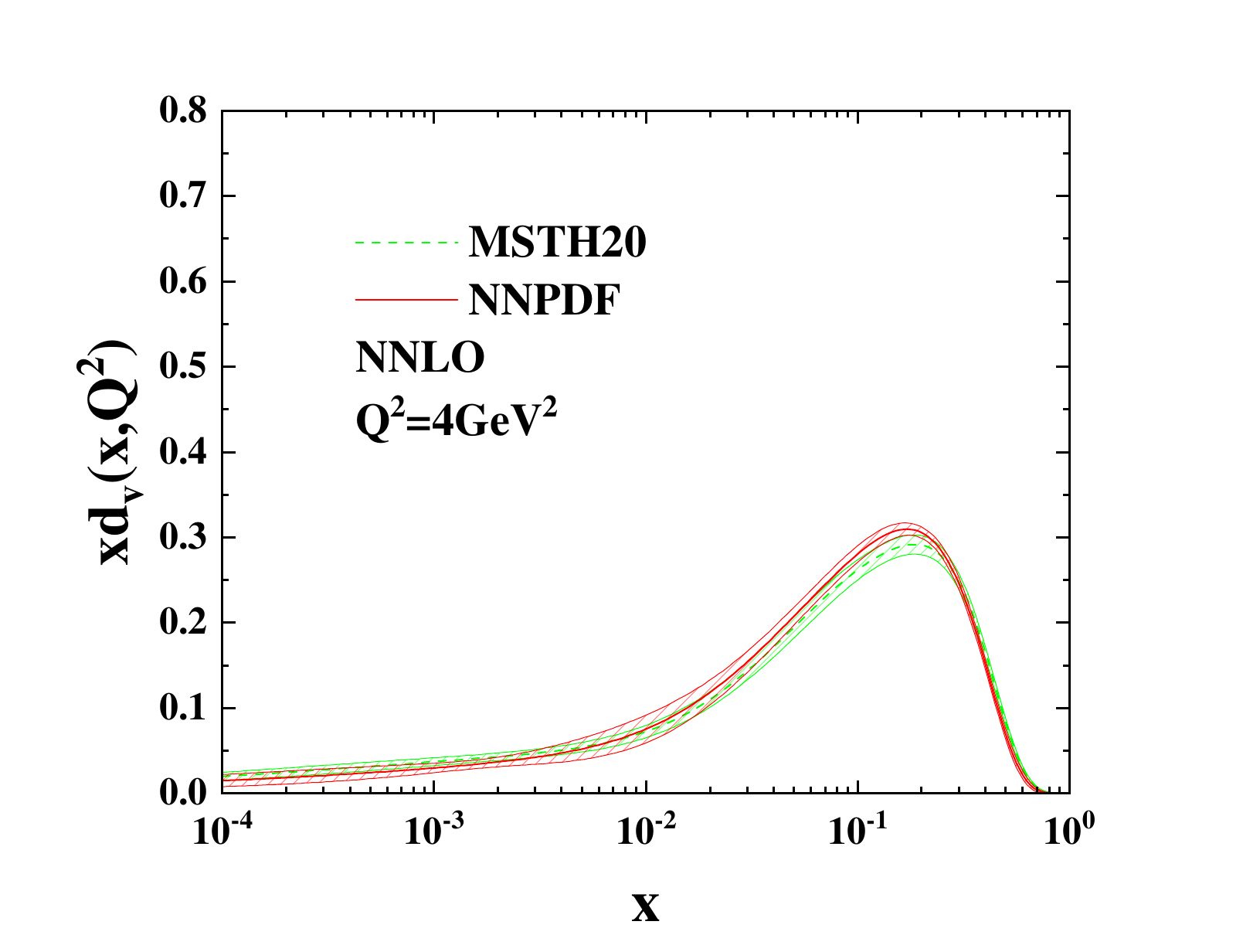 }
	\vspace*{3mm}\\
		\caption{\footnotesize Valence distribution for NNPDF~\cite{NNPDF:2021njg} and MSTH20~\cite{Bailey:2020ooq} in $Q_0^2=4 GeV^2$ .}
	\label{fig:uvdv}
\end{figure*}

	\begin{figure*}
		\includegraphics[clip,width=0.45\textwidth]{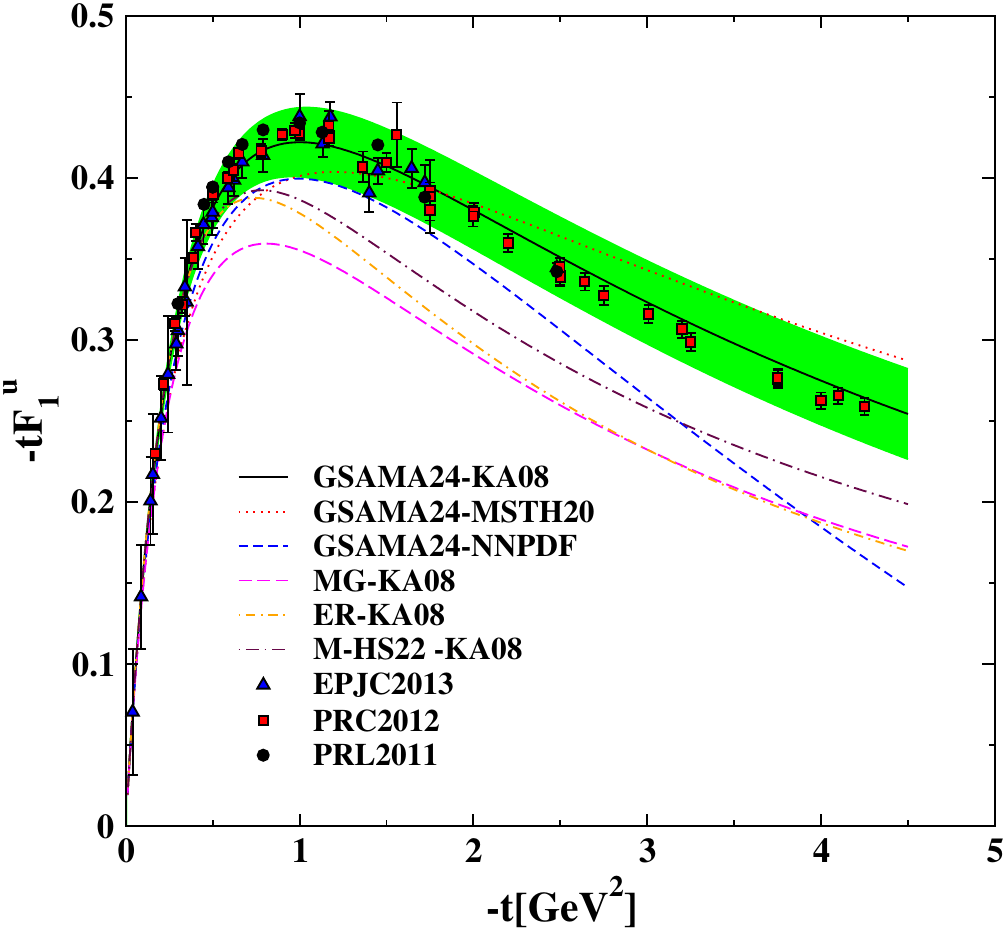 }
							\hspace*{2mm}
		\includegraphics[clip,width=0.45\textwidth]{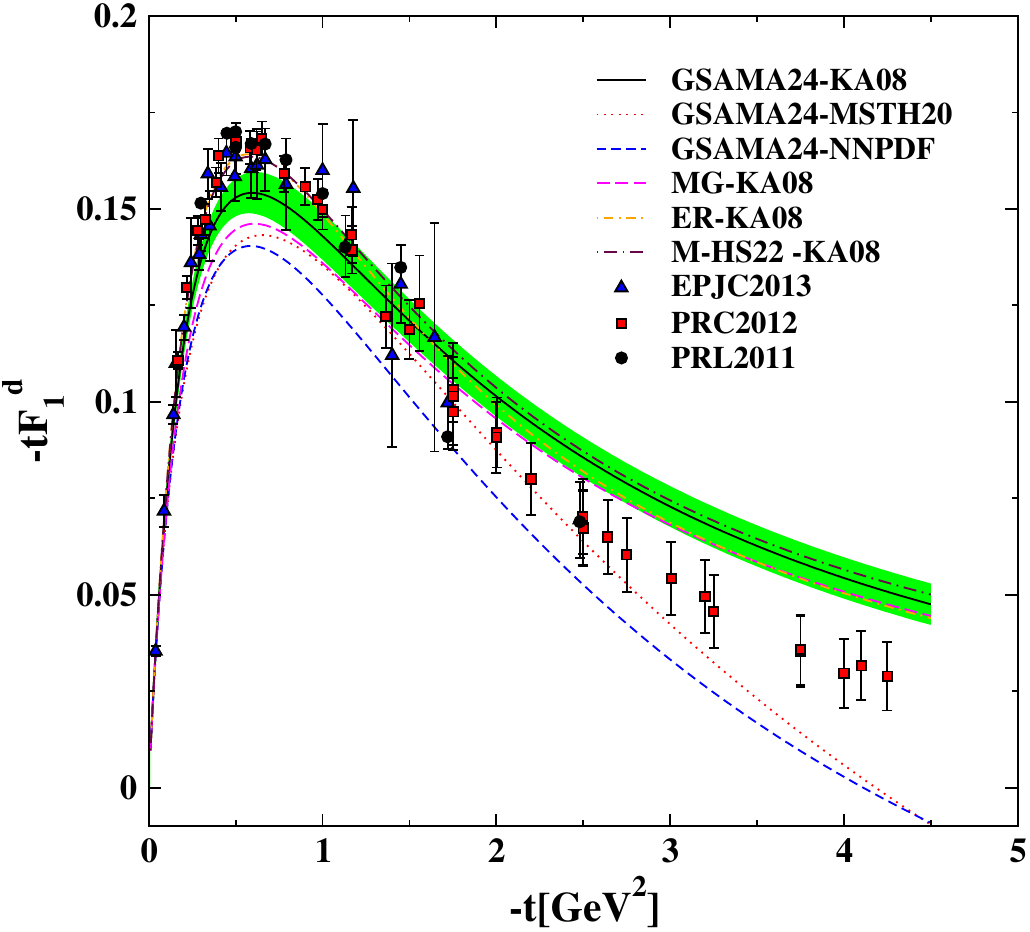 }
									\vspace*{3mm}\\
		\includegraphics[clip,width=0.45\textwidth]{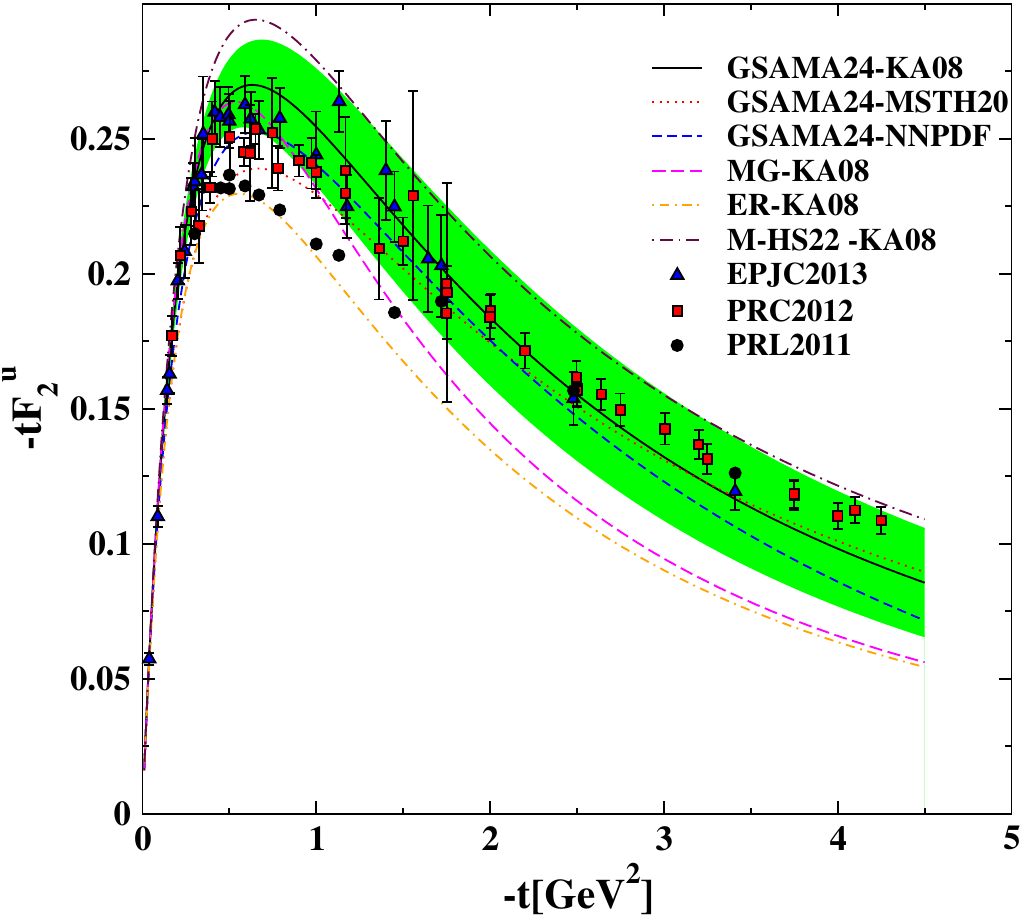 }
									\hspace*{2mm}
		\includegraphics[clip,width=0.45\textwidth]{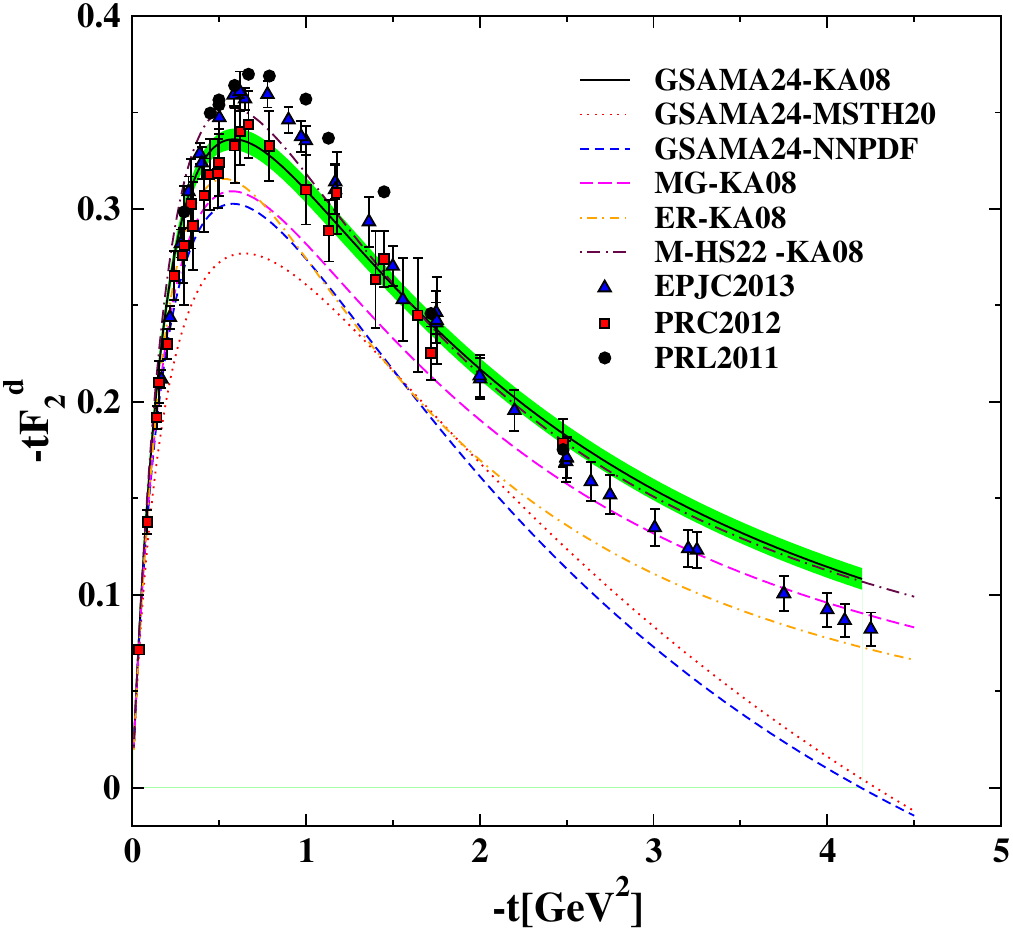 }
			\vspace*{1.5mm}

\caption{\footnotesize The form factors of \(u\) and \(d\) quarks multiplied by \( t \) as a function of \( -t \) using the GSAMA24 ansatz and KA08 ~\cite{Khorramian:2008yh},MSTH20~\cite{Bailey:2020ooq} , NNPDF~\cite{NNPDF:2021njg} PDF. The points shown are extractions based on experimental data from~\cite{Qattan:2012zf,Cates:2011pz,Diehl:2013xca}. Also, the results of the ER model~\cite{Guidal:2004nd}, MG model~\cite{Selyugin:2009ic}, and M-HS22 model~\cite{Vaziri:2023xee} are shown in comparison with the GSAMA24 model.}
	\label{fig:tfud1}
	\end{figure*}
	
	\begin{figure*}
		\includegraphics[clip,width=0.45\textwidth]{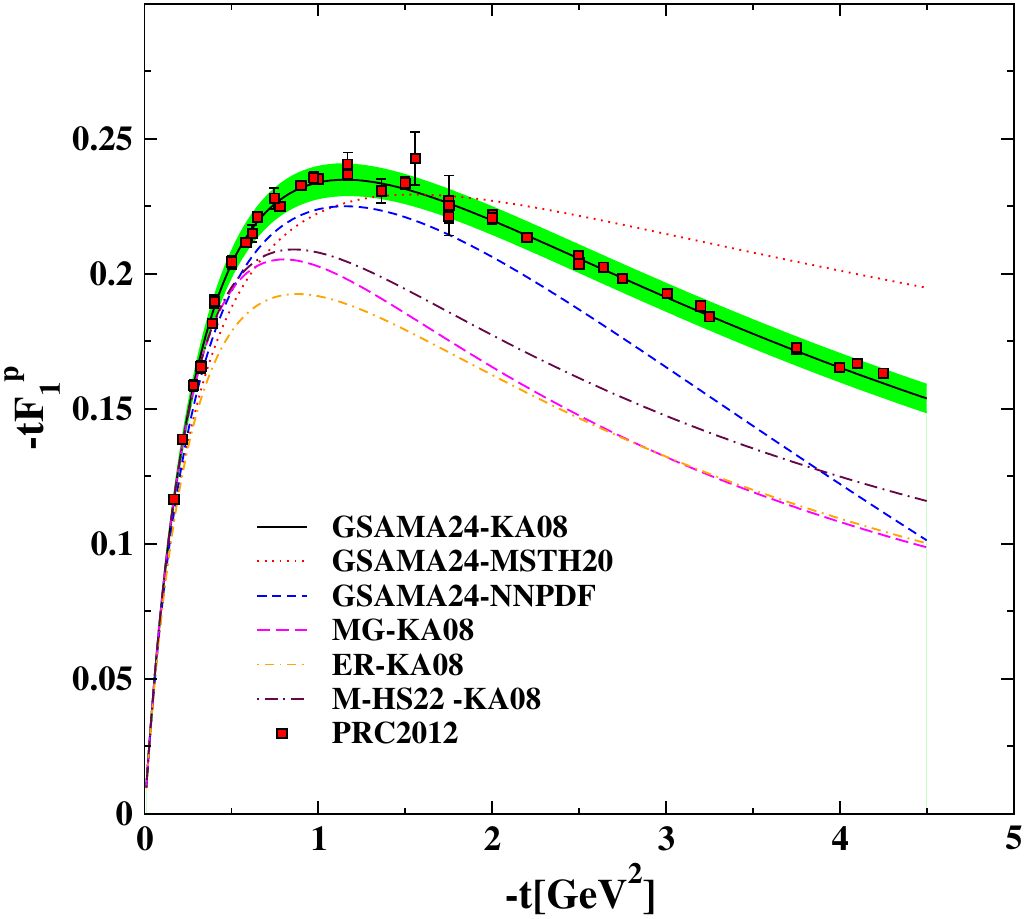 }
									\hspace*{2mm}
		\includegraphics[clip,width=0.45\textwidth]{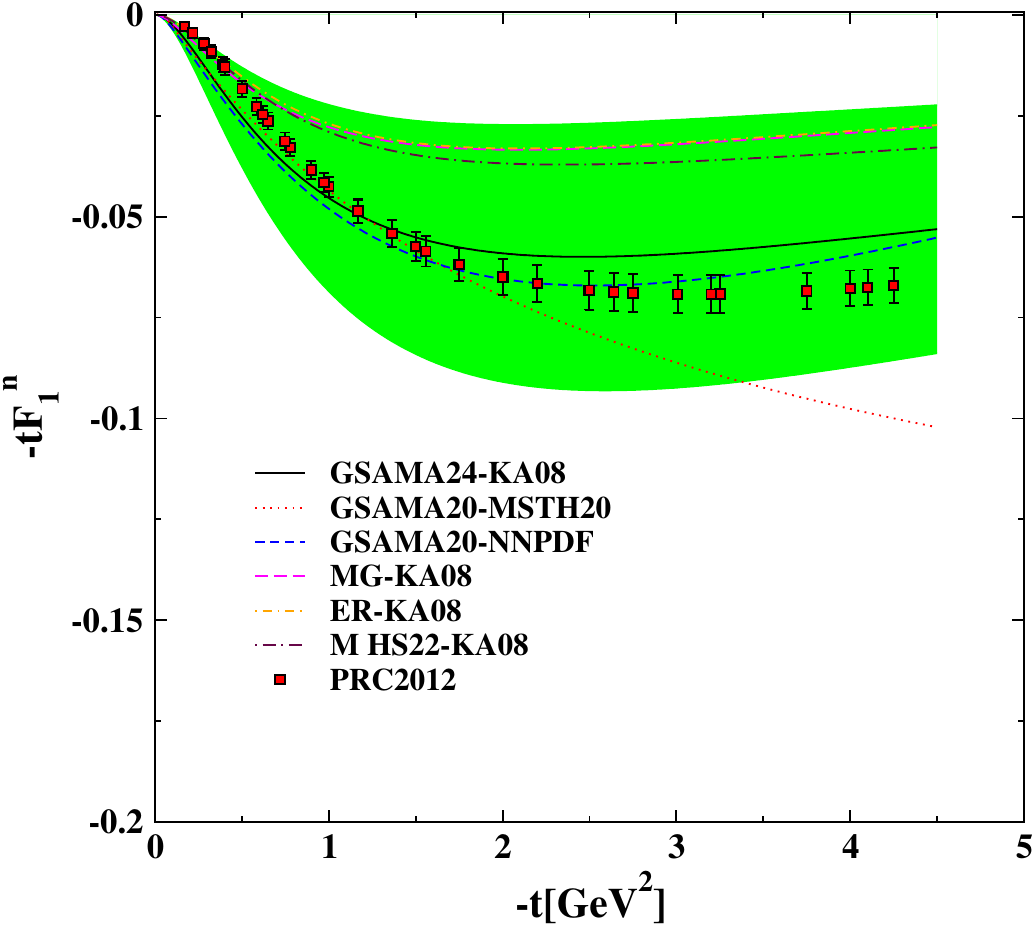 }
											\vspace*{3mm}\\
		\includegraphics[clip,width=0.45\textwidth]{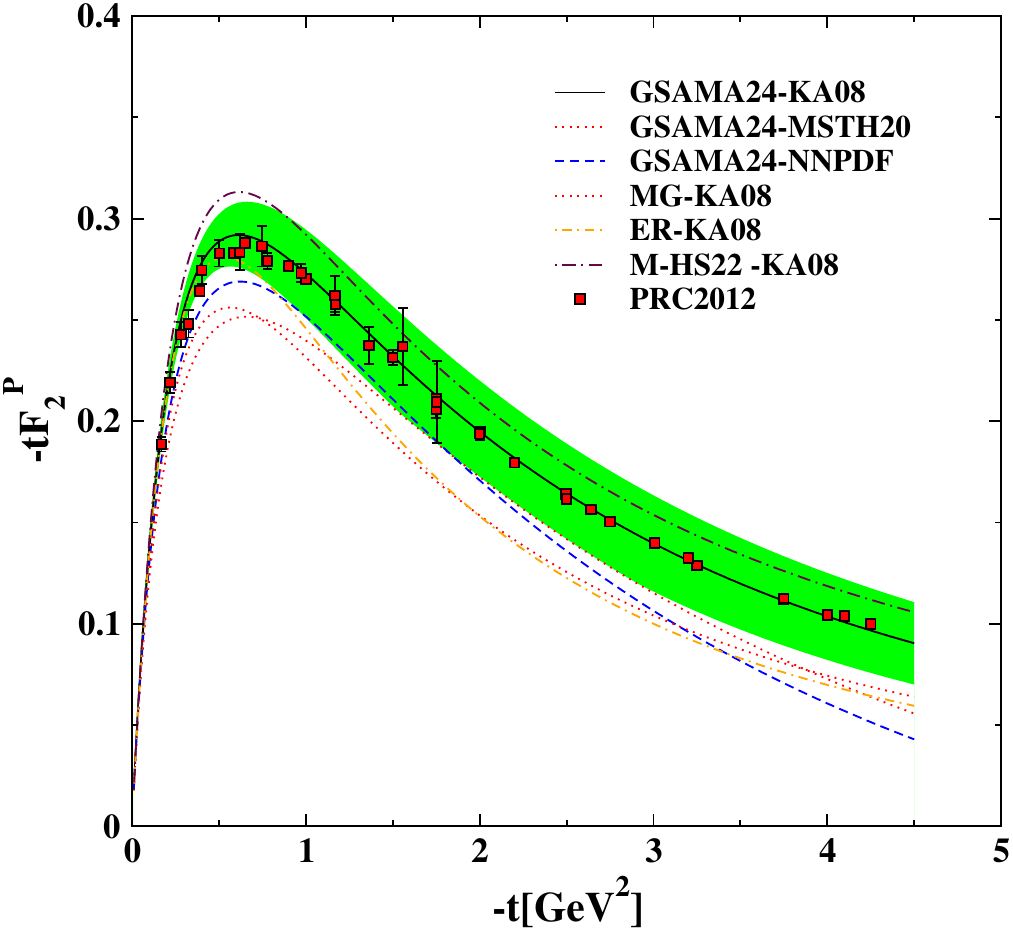 }
									\hspace*{2mm}
		\includegraphics[clip,width=0.45\textwidth]{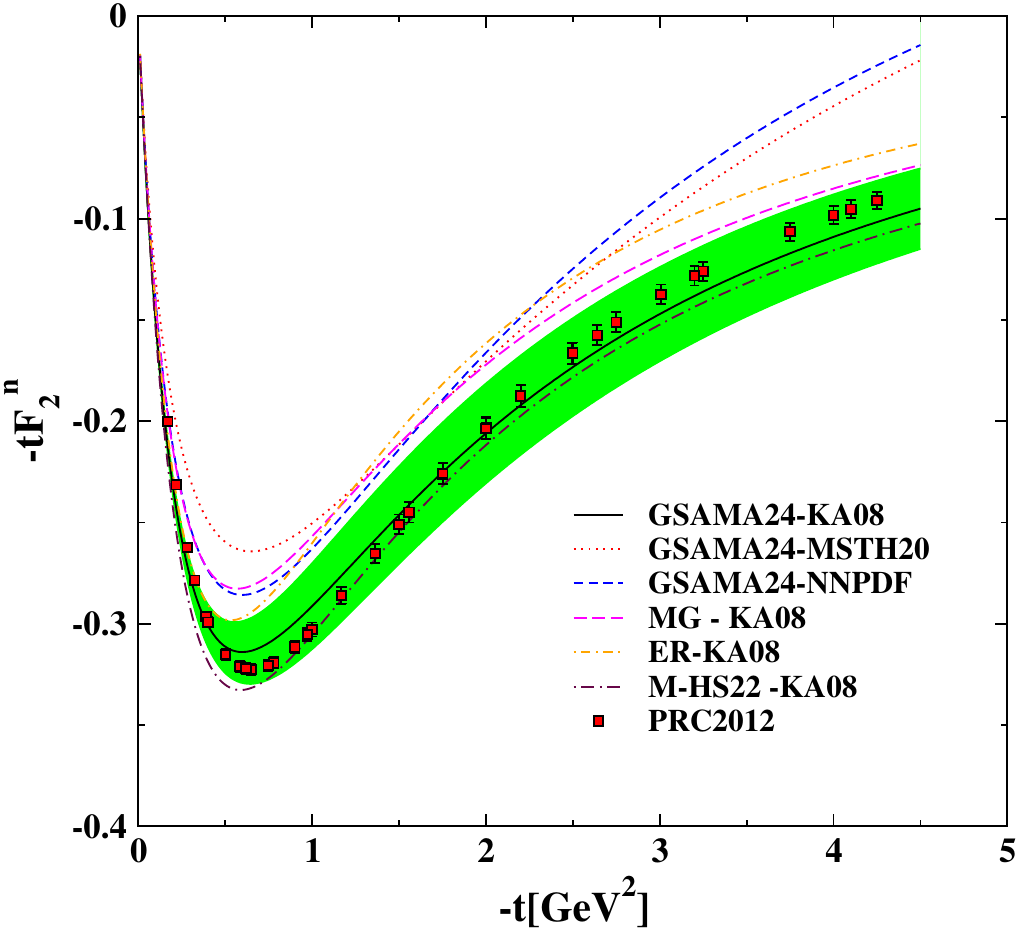 }
									\vspace*{1.5mm}
		\caption{\footnotesize The form factors of proton and neutron  multiplied by $t$ as a function of $-t$ using the GSAMA24 ansatz and KA08~\cite{Khorramian:2008yh}, MSTH20~\cite{Bailey:2020ooq} , NNPDF~\cite{NNPDF:2021njg} PDF. The points shown are extractions based on experimental data from ~\cite{Qattan:2012zf} (square). Also the results of ER model\cite{Guidal:2004nd}, MG model\cite{Selyugin:2009ic} and M-HS22 model\cite{Vaziri:2023xee} are shown in comparison with GSAMA24 model.}
		\label{fig:tfud}
	\end{figure*}

	\begin{figure*}
		\includegraphics[clip,width=0.45\textwidth]{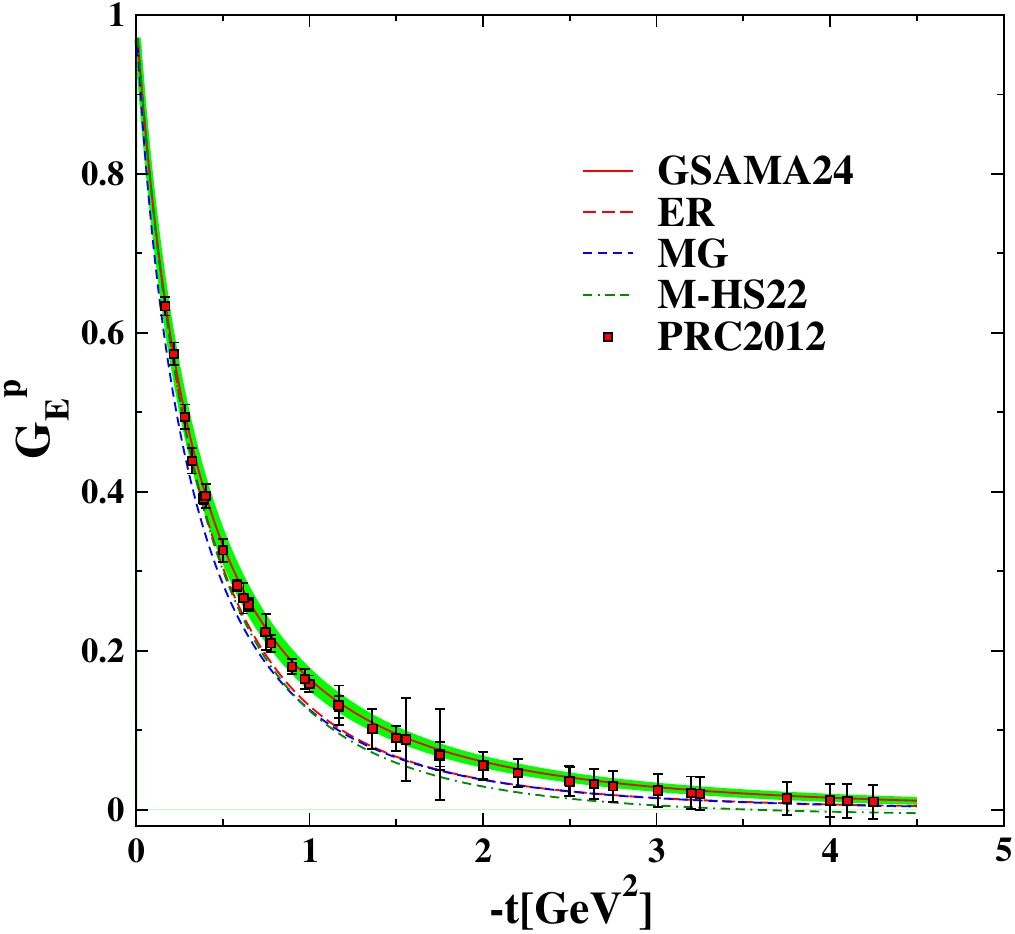 }
													\hspace*{2mm}
		\includegraphics[clip,width=0.45\textwidth]{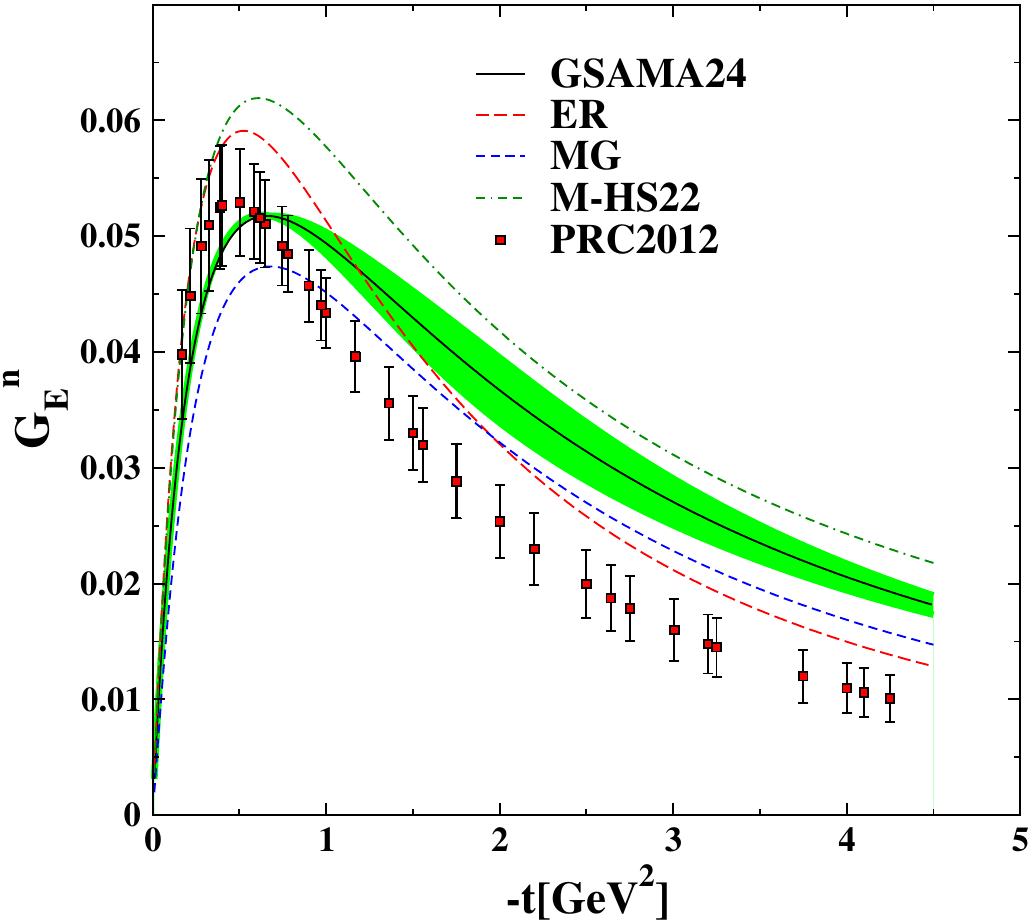 }
											\vspace*{3mm}\\
		\includegraphics[clip,width=0.45\textwidth]{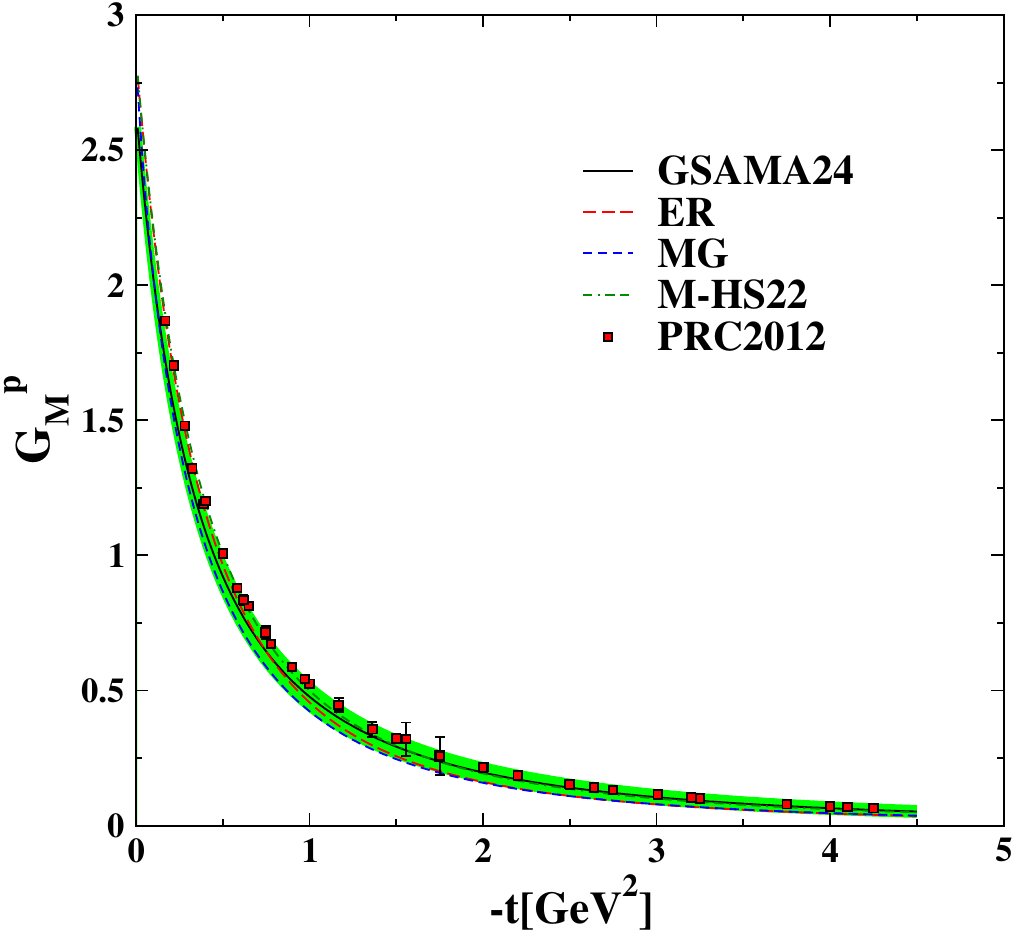 }
															\hspace*{2mm}
		\includegraphics[clip,width=0.45\textwidth]{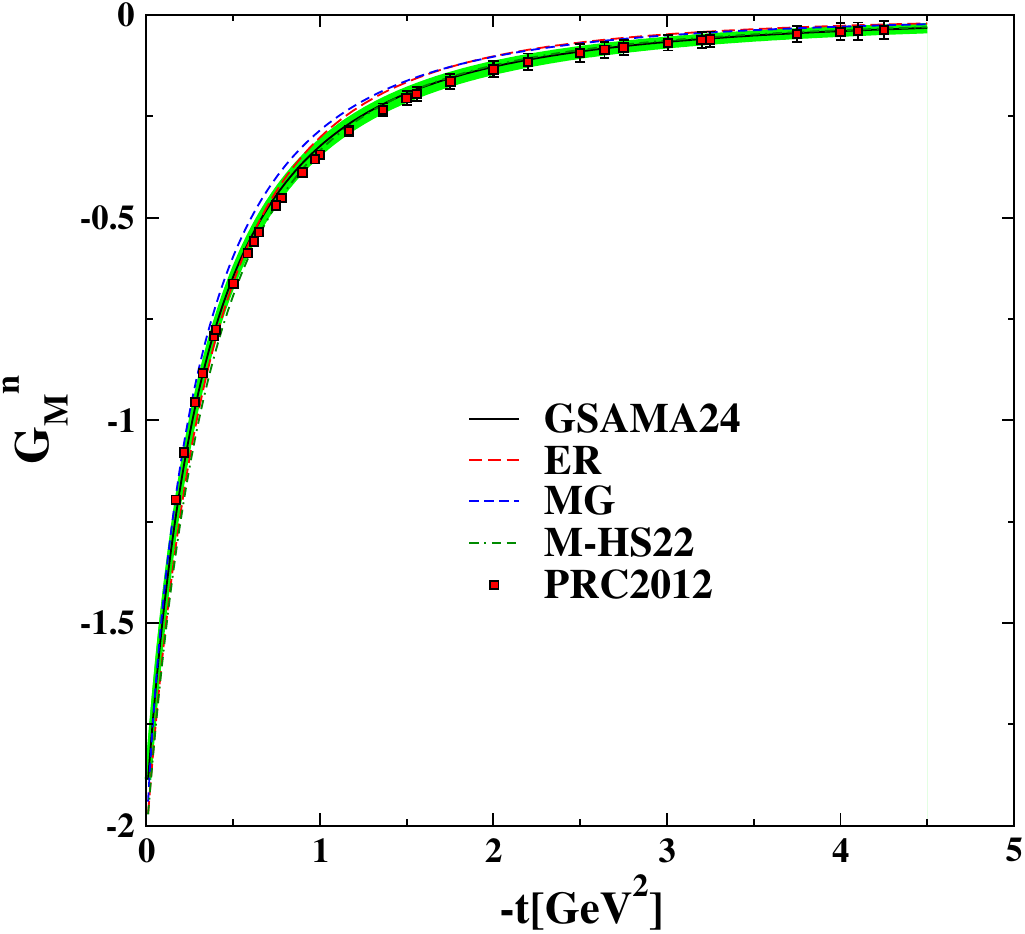 }
			\vspace*{1.5mm}
	\caption{\footnotesize The electric and magnetic form factors \( G_E^{p,n} \) and \( G_M^{p,n} \) as functions of \( -t \). The combination of the M-HS22 model~\cite{Vaziri:2023xee} is compared with the MG model~\cite{Selyugin:2009ic}, and the model~\cite{Guidal:2004nd}. The points shown are extractions based on experimental data from~\cite{Qattan:2012zf} (square).}
	\label{fig:GEMpn}
	\end{figure*}
	
\section{GSAMA24 Ansatz and Nucleon Form Factors}\label{sec3}
Different models exist to calculate the form factors of nucleons, introducing various approximations for more accurate calculations. Among these, we focus on examining three models: ER~\cite{Guidal:2004nd}, MG~\cite{Selyugin:2009ic}, and M-HS22~\cite{Vaziri:2023xee}, and comparing them with our own approximation.

We first explore the MG ansatz, which has been parameterized as follows\cite{Selyugin:2009ic,Selyugin:2014sca,Sharma:2016cnf}:
\begin{equation}
\mathcal{H}^{q}(x,q^2)=q_{v}(x)\exp \left[ \alpha \frac{(1-x)^{2}}{x^{m}}q^2 \right].
\label{eq:HMG}
\end{equation}
\begin{equation}
\varepsilon^{q}(x,q^2)=\varepsilon^{q}(x)\exp \left[ \alpha \frac{(1-x)^{2}}{x^{m}}q^2 \right].
\label{eq:EMG}
\end{equation}
This ansatz has four free parameters which are determined by comparing with experimental data. It includes the parameters \(\alpha=1.15\) and \(m=0.45\) and variable $q$ is defines $u$ and $d$ quarks.

A large portion of modern high-energy scattering experiments, such as proton-proton (or proton-antiproton) collisions and deep-inelastic scattering, relies on the understanding of the partonic constituents of the hadron, known as parton distribution functions (PDFs). By incorporating the 
$q2$-dependence into PDFs, we obtain the 3-D structure of hadrons in terms of generalized parton distributions (GPDs). The Fourier transform of GPDs provides insight into the distribution of charge and magnetization densities of quarks in impact parameter space. Since the $q2$-dependence cannot be derived from first principles, it is valuable to develop a simple parametrization that accurately represents the data on form factors (FFs) over a wide range of momentum transfer.

There are various phenomenological parametrizations for the extraction of GPDs in the literature \cite{Kelly:2004hm}. The simplest model for parametrizing proton GPDs assumes a Gaussian form of the wavefunction, incorporating an interplay between the 
$x$ and $q^2$-dependences \cite{Guidal:2004nd}. Additionally, a Regge parametrization for GPDs, $\mathcal{H}(x,q) =q(x)  \exp[- \alpha\, q^2] $, is applied at small momentum transfers \cite{Goeke:2001tz}, and a modified version of the profile function is often used to further refine the model $q(x) \exp[- \alpha' (1-x) \, q^2]$ is used to extend the analysis to the large momentum transfer region \cite{Guidal:2004nd}. Some of the other well-known methods are based on polynomial or logarithmic forms of the profile function, in line with theoretical and phenomenological constraints \cite{Burkardt:2004bv,Goloskokov:2005sd}. These approaches provide a satisfactory description of the fundamental features of the proton and neutron electromagnetic form factors (EFFs) data.

In a recent parametrization method (PM) \cite{Selyugin:2009ic}, the $q^2$
-dependence is incorporated by combining the Gaussian ansatz with the information on PDFs obtained from global fits to experimental data. However, these calculations are performed using the older set of PDFs from the MRST2002 fit \cite{Martin:2002dr}.
GPD for Dirac form factor for the case of proton
\begin{equation} 
	 \mathcal{H}^p (x,q^2)= \sum_q e_q \, q(x) \, \exp \left[{-a_q^p {(1-x)^2  \over x^{m_p} } q^2} \right] \,, \label{hp}
\end{equation}
where $e_q={2 / 3 }$ for up and $-{1/ 3}$ for down quark and  $a^p_u$, $a^p_d$ and $m_p$ are the free parameters to be fitted from the low $q^2$ experimental data on the proton form factors.
We parametrize the GPD $\mathcal{E}^p(x,q^2)$ for the proton using the widely used representation  \cite{Guidal:2004nd}:
 \begin{equation} 
 	\mathcal{E}^p (x,q^2)= \sum_q e_q  {\cal E}_q (x) \, \exp \left[{-a^p_q { (1-x)^2  \over  x^{m_p} } q^2 } \right] \,, 
 \end{equation} 
 with 
 \begin{eqnarray}  {\cal  E}_u(x) = {\kappa_u \over N_u} (1-x)^{\kappa_1} \, u(x)\,,\\ {\cal E}_d(x) = {\kappa_d \over N_d} (1-x)^{\kappa_2} \, d(x)\,,
 \end{eqnarray}
 where the normalization of up and down quark GPDs to their corresponding anomalous magnetic moments $\kappa_u = 1.673$, $\kappa_d= -2.033$ lead to the $k_1=1.53$, $k_2=0.31$, $N_u =1.52$ and $N_d = 0.95$. We follow the similar kind of parametrization for neutron GPDs while invoking the charge symmetry  and changing the index  $p$ to $n$ for free parameters.

The second ansatz we will use for comparison is the ER model, which consists of three free parameters and is configured as follows\cite{Guidal:2004nd}:
\begin{equation}
\mathcal{H}^{q}(x,t)=q_{v}(x)x^{-\alpha^{\prime}(1-x)t}.
\label{eq:HER}
\end{equation}
\begin{equation}
\varepsilon_{q}(x,t)=\varepsilon_{q}(x)x^{-\alpha^{\prime}(1-x)t}.
\end{equation}
By fitting the experimental data and the form factor of this analysis, we can obtain the value of \(\alpha^{\prime}=1.105\). 
	
The Regge model suggests a behavior of $x^{-\alpha (t)}$ at small 
$x$, or the model:
\begin{eqnarray}
	{\cal H}^q (x,t) = q_v(x)  x^{-(\alpha (t)- \alpha (0))}  
\end{eqnarray}

for the nonforward parton densities ${\cal H}^q (x,t)$. Assuming a linear Regge trajectory with a slope $\alpha^{\, '}$, we get  
\begin{eqnarray}
	{\cal H}^q_{R1} (x,t) = q_v(x) \  x^{- \alpha^{\, '} \, t}  \  .
	\label{eq:hr1}
\end{eqnarray}
This ansatz was already discussed in Ref.~\cite{Goeke:2001tz}. The 
$u$ and $d$ flavor components of the Dirac form factor are given by:
\begin{eqnarray}
	F_1^u(t) \,&=&\, \int _{0}^{1}dx \; u_v(x) \;
	{e^{-t\, \alpha^{\, '}  \ln x} } \nonumber\\
	F_1^d(t) \,&=&\, \int _{0}^{1}dx \; d_v(x) \; 
	{e^{-t \, \alpha^{\, '}  \ln x}}  \  . 
	\label{eq:f1_1}
	\end{eqnarray}
The proton and neutron Dirac form factors follow from:
\begin{eqnarray}
	F_1^p(t) \,=\, e_u \, F_1^u(t) \;+\; e_d \, F_1^d(t) \, , 
	\label{eq:f1p} \\
	F_1^n(t) \,=\, e_u \, F_1^d(t) \;+\; e_d \, F_1^u(t) \, . 
	\label{eq:f1n} 
\end{eqnarray}
By construction, $F_1^p(0)$ = 1, and $F_1^n(0)$ = 0. The Dirac mean squared radii of the proton and neutron in this model are given by:
\begin{eqnarray}
	r^2_{1, p} &\,=\,& -6 \, \alpha^{\, '} \,
	\int _{0}^{1}dx \; 
	\biggl\{ e_u \, \, u_v(x) \,+\, e_d \, \, d_v(x) \biggr\} \, \ln x \; ,
	\label{eq:rms1p}\nonumber \\
	r^2_{1, n} &\,=\,& -6 \, \alpha^{\, '} \,
	\int _{0}^{1}dx \; 
	\biggl\{ e_u \, \, d_v(x) \,+\, e_d \, \, u_v(x) \biggr\} \, \ln x  \;.\nonumber\\
	\label{eq:rms1n}
\end{eqnarray}
Instead of the $1/x$ factor present in the Gaussian model, we now have a much softer logarithmic singularity at small $x$, and the integrals for $r^2_{1}$ converge.
To calculate  $F_2$, we need an ansatz for the nonforward parton densities ${\cal E}^q (x,t)$ . We assume the same Regge-type structure:
\begin{eqnarray}
	{\cal E}^q_{R1}  (x,t) = {\cal E}^q (x)\,  x^{-\alpha^{\, '} t}
\end{eqnarray}
as for ${\cal H}^q (x,t)$. The next step is to model the forward magnetic densities 
${\cal E}^q (x)$. The simplest idea is to take them proportional to the 
${\cal H}^q (x)$ densities. Choosing:
\begin{eqnarray}
	{\cal E}^u (x) = \frac{\kappa_u}{2} u_v(x)  
	\qquad {\rm and}  \qquad {\cal E}^d (x) = \kappa_d d_v(x) \  ,
	\label{eq:er1}
\end{eqnarray}
we satisfy the normalization conditions that guarantee 
$F_2^p(0)$ = $\kappa_p$ and $F_2^n(0)$ = $\kappa_n$.
the Regge model $R1$ fits $F_1^p(t)$ and 
$F_2^p(t)$ data for small momentum transfers 
$-t \lesssim 0.5$\,GeV$^2$. However, the suppression at larger 
$-t$ in the $R1$ model is too strong, and it falls considerably short of the data for 
$-t > 1$\,GeV$^2$ .

To improve agreement with the data at large 
$-t$, we need to modify our models. Both the Gaussian (G) and the Regge-type model (R1) discussed earlier have the structure:
$${\cal H}(x,t) = 
q_v(x) \exp[\,tg(x)] \ ,
$$ 

with $g(x) \sim (1-x)/x$ for the Gaussian model and $g(x) \sim -\ln x$ for the Regge-type model. Thus, at large $t$, the form factors are dominated by integration over regions where $tg(x) \sim 1$ or $g(x) \sim 1/t \to 0$. In both models, $g(x)$ approaches zero only as $x \to 1$,meaning that the large-
$t$ asymptotics of $F_i(t)$ is determined by the $x \to 1$ region. Given that $g(x)\sim 1-x$ as  $x \to 1$, if  $q_v(x)\sim (1-x)^{\nu}$ for $x$ near 1, then the form factors will decrease as $1/t^{\nu +1}$ at large $t$. Experimentally, $\nu$ is close to 3, so the G and R1 models predict $a \sim 1/t^4$ behavior for the form factors. This appears to contradict the experimentally observed $1/t^2$  behavior of $F_1^p (t)$ , leading to the potential conclusion that these models may fail to describe the data.However, it is important to note that the model curves for $F_1^p (t)$ are more complex than just a simple power law $\sim~1/t^4$. In fact, up to 10\,GeV$^2$, the Gaussian model reproduces the data for $F_1^p$ within 10\%~\cite{Radyushkin:1998rt}.For higher $t$ , the Gaussian model prediction for $F_1^p$ decreases faster than $1/t^2$ and falls below the data. However, the nominal $1/t^4$ asymptotic behavior is only reached at extremely large values of $-t\sim$ 500\,GeV$^2$.Thus, the simplest approach is to introduce an extra $(1-x)$ factor into the original $g(x)$ functions. To preserve the Regge structure at small $x$ and $t$, we propose the modified Regge ansatz R2~\cite{Burkardt:2002hr,Burkardt:2004bv}:
\begin{eqnarray}
	{\cal H}^q_{R2} (x,t) = q_v(x)  x^{-\alpha^{\, '} (1-x)t}  \  .
	\label{eq:hr2}
\end{eqnarray}

And the last ansatz is the M-HS22~\cite{Vaziri:2023xee}, which has been published recently as the modified mode of the HS22 ansatz and is presented as\cite{Vaziri:2023xee}:
\begin{equation}
\mathcal{H}^{q}(x,t)=q_{v}(x)\exp [-\alpha^{\prime \prime} t(1-x)\ln(x)+\beta x\ln(1-bt)],
\label{eq:H}
\end{equation}
\begin{equation}
\varepsilon^{q}(x,t)=\varepsilon^{q}(x)\exp [-\alpha^{\prime \prime} t(1-x)\ln(x)+\beta x\ln(1-bt)],
\label{eq:E}
\end{equation}
We introduce changes to the parameters \(\alpha^{\prime \prime}\), \(\beta\), and \(b\), which are taken to be 1.125, 0.185 and 2, respectively, for M-HS22~\cite{Vaziri:2023xee}. 

Each of the three introduced approaches provides a somewhat suitable description of nucleon form factors. However, it is evident that these calculations have a lot of room for improvement, and new approaches need to be introduced to provide better, more suitable, and more compatible results with experimental data. Therefore, the GSAMA24 group has initiated new calculations to find a more appropriate description for the form factors. The parameterization we suggested for this analysis is as follows:
\begin{equation}\label{eq:HGS}
\mathcal{H}^{q}(x,t)=q_{v}(x)\exp[c t(1-x)^g\ln(x)+e x^{b}\ln(1+dt)]
\end{equation}
\begin{equation}\label{eq:eGS}
\varepsilon^{q}(x,t)=\varepsilon_{q}(x)\exp[c t(1-x)^g\ln(x)+e x^{b}\ln(1+dt)]
\end{equation}
We applied the GSAMA24 ansatz to experimental form factor data using a least-squares fitting method to minimize the discrepancies between the predicted and observed values. A brief overview of the used dataset is provided in Table \ref{tab:tab11}, highlighting its significance as a comprehensive source of experimental measurements. We calculate the $ \chi^2/{d.o.f} $ value to quantify the quality of our fit.
 By fitting the calculations with experimental data, the coefficients of Eq.\ref{eq:HGS} and Eq.\ref{eq:eGS} were found as shown in Table \ref{tab:tab1}. The needed \(\eta_{u}\) and \(\eta_{d}\) for GSAMA24 and other introduced modelss are presented in Table \ref{tab:eta}. The values of $\eta_{u,d}$  are sensitive to the theoretical framework and fit methodologies employed in their derivation. Variations in the input data can cause corresponding shifts in the  $\eta_{u,d}$ factors, reflecting the complexities involved in accurately representing the internal dynamics of quarks.
It is essential to interpret the  $\eta_{u,d}$  factors not just as numerical values but also in the context of the interactions they represent. Disparities in these factors could signify underlying differences in the quark distributions and how they respond to the applied conditions, such as energy levels or collision dynamics in high-energy physics.
Our results highlight the inherent complexity of quark dynamics and interactions within nucleons. While large variations might initially seem alarming, they can also be indicative of the diverse conditions under which these factors are derived, reflecting the multifaceted nature of particle interactions.
 We also calculated the $\chi^2/{d.o.f}$ values for all models included in the analysis, as summarized in Table \ref{tab:tab3}. By incorporating these values, we allow for direct evaluation of the models' performances against the same set of experimental measurements. A lower value of $\chi^2/{d.o.f}$ indicates a better fit, enabling straightforward comparisons among the various models. To enhance our analysis further, we dissected the $\chi^2/{d.o.f}$ results into components based on separate experimental contributions, specifically separating the up/down quark and proton/neutron datasets. This separation provides a more nuanced view of the models' performances and offers insights into how well each model captures the specific characteristics of different quark distributions. 

\begin{table}
		\caption{The data used for analysis.}
	\label{tab:tab11}
\begin{tabular}{ccc}
	\multicolumn{3}{c}{Data} \\\hline 
	$tF_{1}^{u}$ & \cite{Qattan:2012zf,Cates:2011pz,Diehl:2013xca}&PRC2012,PRL2011,EPJC2013 \\ 
	$tF_{1}^{d}$ & \cite{Qattan:2012zf,Cates:2011pz,Diehl:2013xca}&PRC2012,PRL2011,EPJC2013 \\ 
	$tF_{1}^{n}$ & \cite{Qattan:2012zf}&PRC2012 \\ 
	$tF_{1}^{p}$ &  \cite{Qattan:2012zf}&PRC2012\\ 
	$tF_{2}^{u}$ & \cite{Qattan:2012zf,Cates:2011pz,Diehl:2013xca}&PRC2012,PRL2011,EPJC2013 \\ 
	$tF_{2}^{d}$ & \cite{Qattan:2012zf,Cates:2011pz,Diehl:2013xca}&PRC2012,PRL2011,EPJC2013 \\ 
	$tF_{2}^{n}$ & \cite{Qattan:2012zf}&PRC2012 \\ 
	$tF_{2}^{p}$ & \cite{Qattan:2012zf}&PRC2012\\\hline
\end{tabular}
\end{table}

\begin{table*}[htb]
\caption{The values of free parameters of the GSAMA24 ansatz in Eqs.\ref{eq:HGS},\ref{eq:eGS}.}
\label{tab:tab1}
\begin{tabular}{lllll}
\hline
\(c=1.8809\pm 8.7925\times 10^{-3}\) & \(d=1.26758\pm0.1596\)&\(b=7.0492\times 10^{-2}\pm7.594\times 10^{-3}\) & \\
\(e =1.6978\pm 0.1966\) & \(g=0.21681\pm7.0602\times10^{-2}\) &\(\chi^2/d.o.f=1314.4925/386=3.4054\)& \\
\hline
\end{tabular}
\end{table*}

\begin{table}[h]
\caption{Values of $\eta_{u}$ and $\eta_{d}$ in the GSAMA24 and other three models.}
\label{tab:eta}
\begin{tabular}{ccc}\hline
Model & $\eta_{u}$ & $\eta_{d}$ \\\hline
GSAMA24 &1.0338 & 0.078971 \\
ER~\cite{Guidal:2004nd} & 1.713 & 0.566 \\
MG~\cite{Selyugin:2009ic} & 1.8 & 0.31 \\
M-HS22~\cite{Vaziri:2023xee} & 0.7 & 0.19\\\hline
\end{tabular}
\end{table}
\begin{table*}[htb]
	\caption{The values of $\chi ^{2}/d.o.f$ of the GSAMA24 ansatz and ER, MG,M-HS22 for data analysis .}
	\label{tab:tab3}
\begin{tabular}{ccccc}
	\hline
	\multicolumn{5}{c}{$\chi ^{2}/d.o.f$} \\ \hline
	Model & ER & MG & M-HS22 & GSAMA24 \\ \hline\hline
	$tF_{1}^{u}$ & 8797.878/66=133.301 & 15819.314/66=239.686 & 
	272220.653/66=4124.555 & 115.4061/66=1.74857$\pm3.28\times10^{-6}$ \\ 
	$tF_{1}^{d}$ & 67.611/66=1.0244 & 797.434/66=12.082 & 35764.706/66=541.889 & 
	211.3657/66=3.20251$\pm1.17\times10^{-6}$ \\ 
	$tF_{1}^{n}$ & 1623.347/39=41.624 & 1690.590/39=43.348 & 7460.836/39=191.303
	& 192.6823/39=4.94057$\pm1.89\times10^{-6}$\\ 
	$tF_{1}^{p}$ & 34165.117/39=876.028 & 49188.051/39=1261.232 & 
	566016.797/39=14513.251 & 55.79048/39=1.43052$\pm5.19\times10^{-6}$ \\ 
	$tF_{2}^{u}$ & 1235.057/39=31.668 & 1023.829/39=26.252 & 25829.994/39=662.307
	& 165.2981/39=4.23841$\pm4.32\times10^{-6}$ \\ 
	$tF_{2}^{d}$ & 496.660/39=7.525 & 5450.252/39=82.579 & 50187.171/39=760.411
	& 203.6131/66=3.08504$\pm7.91\times10^{-6}$ \\ 
	$tF_{2}^{n}$ & 1957.534572/39=293.605 & 463.044/39=11.872 & 
	257506.702/39=6602.735 & 251.8093/39=6.45665$\pm2.26\times10^{-6}$ \\ 
	$tF_{2}^{p}$ & 11450.605/39=293.605 & 9663.730/39=247.787 & 
	245894.272/39=6304.981 & 118.5273/39=3.03916$\pm2.52\times10^{-6}$ \\ \hline\hline
\end{tabular}
\end{table*}

Nevertheless, we use the KA08 PDF~\cite{Khorramian:2008yh} for analysis.
The KA08 parton distribution functions in the NNLO approximation are discussed below for a range of input $Q_0^2 = 4.0 \text{ GeV$^2$}$~\cite{Khorramian:2008yh}: 
\begin{equation}
xu_v=3.1357719x^{0.7858}(1 - x)^{3.6336}(1 + 0.1838 x^{0.5} - 1.2152 x)\label{eq:xuv}
\end{equation}
\begin{equation}
xd_v=4.868494x^{0.7772}(1 - x)^{4.0034}(1 + 0.1 x^{0.5} + 1.14 x)\label{eq:xdv}
\end{equation}
Additionally, the behavior of newer PDFs such as MSTH20~\cite{Bailey:2020ooq} and NNPDF~\cite{NNPDF:2021njg} in the NNLO approximation at $Q_0^2 = 4.0 \text{ GeV}^2$ is shown in Fig.~\ref{fig:uvdv}. To test these recent PDF sets, we incorporated MSTH20~\cite{Bailey:2020ooq} and NNPDF~\cite{NNPDF:2021njg} into our ansatz in the NNLO approximation at $Q_0^2 = 4.0 \text{ GeV}^2$ , as illustrated in Fig.~\ref{fig:tfud1} and Fig.~\ref{fig:tfud}. We then compared these with the KA08~\cite{Khorramian:2008yh} parton distribution analyzed in this project. Error bands were calculated using the Hessian method. As observed, KA08 aligns more closely with the experimental data than the other PDFs."\\
Having all the necessary factors to calculate the nucleon form factor as mentioned in Section \ref{sec:sec2}, utilizing the mentioned PDFs, and finally possessing the parametrized ansatz GSAMA24, we can compute the $F_{1}(t)$ function using  $\mathcal{H}(x,t)$ and calculate the $ F_{2}(t)$ function with  $\varepsilon(x,t)$ for the quarks u and d. This allows us to reach the hierarchray of factors related to protons and neutrons. By performing these calculations, we can illustrate the changes in the form factors related to the u and d quarks as well as the proton and neutron in terms of $t$, the results of which are shown in Figs. \ref{fig:tfud1} and \ref{fig:tfud}.

The results show that the ansatz GSAMA24 has good compatibility with experimental data and has provided better results compared to three other ansatze in almost all aspects.

The electric $G_{E}$ and magnetic $G_{M}$ form factors are pivotal in understanding the nucleon's internal structure, as they encapsulate information about the spatial charge and current distributions with in the nucleon \cite{Qattan:2012zf,Ernst:1960zza}.

\begin{equation}
G_{E}^{N}(t) = F_{1}(t) + \frac{t}{4M^{2}}F_{2}(t), \;\;\; G_{M}^{N}(t) = F_{1}(t) + F_{2}(t).
\label{eq:GN}
\end{equation}

These form factors are functions of the squared four-momentum transfer  $t$, and in the limit as $t$ approaches zero, they reveal the nucleon's static properties, such as its charge and magnetic moment. 

The simplicity of the nonrelativistic interpretation lies in its assumption that the nucleon's charge and magnetization are predominantly borne by its constituent quarks, the up and down quarks. These quarks are thought to have similar spatial distributions, which results in their comparable influence on the form factors.
By analyzing the contributions of up and down quarks to the nucleon's form factors, researchers can delve deeper into Quantum Chromodynamics (QCD) that governs the interactions and dynamics of these fundamental particles. Such studies not only enhance our knowledge of nucleon structure but also contribute to the broader field of particle physics, where understanding the most basic units of matter is paramount.
Through ongoing research and experimentation, scientists continue to probe the depth of nucleon structure, seeking to unravel the complexities of the forces at play within atomic nucleons. The insights gained from these endeavors not only enrich our theoretical frameworks but also pave the way for potential technological advancements rooted in the principles of quantum mechanics and nuclear physics \cite{Beck:2001yx}.

\begin{eqnarray}
G^p_{E,M} &=& \frac{2}{3}G^u_{E,M} - \frac{1}{3}G^d_{E,M}, \nonumber \\
G^n_{E,M} &=& \frac{2}{3}G^d_{E,M} - \frac{1}{3}G^u_{E,M}.
\end{eqnarray}

Here is an expanded form of the expression for the up and down quark contributions to the proton form factors~\cite{Miller:1990iz}:

\begin{eqnarray}
G^u_{E,M} &=& G^n_{E,M} + 2G^p_{E,M}, \nonumber \\
G^d_{E,M} &=& 2G^n_{E,M} + G^p_{E,M}.
\end{eqnarray}

In the framework of the standard model, the form factors $G_{E,M}^{u}$ and $G_{E,M}^{d}$ encapsulate the contributions of up and down quarks to the proton's and neutron's electric and magnetic properties. These form factors are pivotal in quantifying the intrinsic electromagnetic characteristics of these subatomic particles. The form factors $F_{1}$ and $F_{2}$ are similarly governed by analogous equations, reflecting the underlying symmetry in the quark contributions to the nucleon structure.\\
The magnetic moments of the quarks, denoted as $\mu_{u}$ and $\mu_{d}$, are inferred from the behavior of the magnetic form factors in the limit as the momentum transfer approaches zero.\\
Precisely, the magnetic moments are given by:

\begin{eqnarray}
\mu_{u} &=& (2\mu_{p} + \mu_{n}) = 3.67\mu_{N}, \\
\mu_{d} &=& (\mu_{p} + 2\mu_{n}) = -1.03\mu_{N}.
\end{eqnarray}

Here,  $\mu_{N}$ represents the nuclear magneton, a physical constant used as a reference unit for the magnetic moment of nucleons. It is important to recognize that the terms "up and down quarks contributions" encompass the net effects of both quarks and antiquarks. This distinction is vital because it reflects the net difference in their contributions within the nucleons, which is a consequence of the charge-weighted sum of the individual contributions to the form factors.

This nuanced understanding is essential for interpreting experimental data and theoretical models accurately. In Figure \ref{fig:GEMpn}, the electric $G_{E}$ and magnetic $G_{M}$ form factors for protons and neutrons are illustrated. These form factors are derived from a synthesis of various ansatz models and proton distribution functions, offering a comprehensive view of the nucleon structure.

The graphical presentation plots these form factors against the negative squared momentum transfer $-t$, providing a visual depiction of how these factors vary with the energy scale of the interaction. The intricate interplay between the quark distributions and the resulting electromagnetic form factors is a subject of ongoing research. By employing different models and distribution functions, researchers aim to refine their understanding of the nucleon's internal structure and the fundamental forces that govern its behavior.It is important to note that the up- and down-quark contributions, as defined in this context, include contributions from both quarks and antiquarks, and they represent the difference between the quark and antiquark distributions due to the charge weighting of their contributions to the form factors~\cite{Perdrisat:2006hj}.

Overall, Our analysis reveals that the GSAMA24 model, when compared with the KA08 parton distribution, often yields results that show the best agreement with experimental data. As illustrated in Figures \ref{fig:tfud1} and \ref{fig:tfud}, the form factors of the $u$ and $d$ quarks, as well as the form factors for the proton and neutron multiplied by $t$, for both the ER and MG models, are largely consistent with each other while diverging from the experimental data from EPJC 2013, PRC 1012, and PRL 2011. Among the models evaluated, GSAMA24 demonstrates the highest level of agreement with the experimental results.
In Figure \ref{fig:GEMpn}, which compares the electric and magnetic form factors to experimental data from PRC 2012, the GSAMA24 model continues to demonstrate the best fit, outperforming the ER, MG, and M-HS22 models for $G^E_n$ , $G^M_p$ , and $G^M_n$. Additionally, for $G^E_n$, GSAMA24 is closer to the data at small values of $x$. 
The theoretical framework used to compute $G^E_n$ relies on specific assumptions regarding the structure of nucleons and quark interaction dynamics, and variations in model parameters or approximations can lead to shifts in predicted values. While the agreement for $G^E_n$ is generally good at small $x$, shifts at larger $x$ may result from the use of proton parton distribution functions, particularly affecting the neutron's $G^E_n$.
To improve the alignment of our results with experimental data for both protons and neutrons, we plan to conduct a simultaneous QCD fit on both ansatze and generalized parton distributions (GPDs).

\section{Dirac Mean Squared Radii}\label{sec4}
In parton distribution functions (PDFs), the Dirac mean-squared radii quantify the size of a proton or neutron, which is based on the spatial distribution of its constituent partons. This measurement is derived from the analysis of form factors. Within PDFs, the Dirac mean-squared radius is linked to the isoscalar electric charge form factor, offering insights into the nucleon's low-energy structure.

The spectral analysis of the mean-squared radii is instrumental in distinguishing the various structural contributions to the nucleon, such as the core and the meson cloud. It also aids in understanding the dynamics of nucleons across different energy scales. The Dirac mean-squared radii are essential for comprehending the intrinsic properties of nucleons and their interactions within the nucleus, as well as in high-energy physics.
\begin{table*}[htb]
	\begin{center}
		\caption{{\footnotesize The electric radii of the proton and neutron were calculated using KA08~\cite{Khorramian:2008yh} parton distribution functions (PDFs) based on the extended (ER)~\cite{Guidal:2004nd}, (MG)~\cite{Selyugin:2009ic} and (M-HS22)~\cite{Vaziri:2023xee} models. The data used in this study are obtained from \cite{ParticleDataGroup:2010dbb}.}
			\label{tab:tabR}}
		\vspace*{1.5mm}
		\begin{tabular}{ccc}
			\hline\hline
			PDFs & $r_{E,p}$ & $r^2_{E,n} $   \\   \hline
			Experimental data   & 0.877$\pm$0.009(stat)$\pm$0.011(syst) $fm$ & -0.115$\pm$0.013(stat)$\pm$0.007(syst) $fm^2$ \\
			GSAMA24-KA08   & 0.861893$\pm$0.061841 $fm$ & -0.116988$\pm$0.0025322 $fm^2$ \\
			ER-KA08    & 0.813667 $fm$ & -0.114973 $fm^2$ \\
			MG-KA08    & 0.983821 $fm$ & -0.112921 $fm^2$ \\
			M-HS22-KA08   & 0.845524 $fm$ & -0.115755 $fm^2$ \\
			\hline\hline					
		\end{tabular}
	\end{center}
\end{table*}
In this approach, we first introduce the relationships and results of the GSAMA24 model, which can calculate the electric radius of the proton and neutron as follows:
\begin{eqnarray}
r_{1,p}^{2} &=& -6 \int_{0}^{1}dx\left[e_{u}u_{v}(x)+e_{d}d_{v}(x)\right]\\ \nonumber
&&\left(de x^{b}+c(1-x)^g\ln(x)\right),\label{eq:GSAMA241}
\end{eqnarray}
\begin{eqnarray}
r_{1,n}^{2} &=& -6 \int_{0}^{1}dx\left[e_{u}d_{v}(x)+e_{d}u_{v}(x)\right]\\ \nonumber
&&\left(de x^{b}+c(1-x)^g\ln(x)\right).\label{eq:GSAMA242}
\end{eqnarray}
In particular, the electric mean squared radii of the proton and neutron are given by
\begin{equation}\label{r1}
r^2_{E,p} = r_{1,p}^2 + \frac{3}{2}\frac{\kappa_p}{M_N^2},
\end{equation}
\begin{equation}\label{r2}
r^2_{E,n} = r_{1,n}^2 + \frac{3}{2}\frac{\kappa_n}{M_N^2},
\end{equation}
where the first term in Equations \ref{r1} and \ref{r2} is the Dirac radius squared \( r_{1}^{2} \), whereas the second term is the Foldy term~\cite{Guidal:2004nd}.

The nucleon's Dirac mean-squared radii based on the extended Regge ansatz (ER)~\cite{Guidal:2004nd} are
\begin{equation}
r_{1,p}^{2} = -6\alpha^{\prime}\int_{0}^{1}dx\left[e_{u}u_{v}(x)+e_{d}d_{v}(x)\right](1-x)\ln(x),\label{eq:rpREEGE}
\end{equation}
\begin{equation}
r_{1,n}^{2} = -6\alpha^{\prime}\int_{0}^{1}dx\left[e_{u}d_{v}(x)+e_{d}u_{v}(x)\right](1-x)\ln(x),\label{eq:rnREEGE}
\end{equation}

Furthermore, the electric radii of the proton and neutron have been determined utilizing the MG model~\cite{Selyugin:2009ic}:
\begin{equation}
r_{1,p}^{2} = 6\alpha \int_{0}^{1}dx\left[e_{u}u_{v}(x)+e_{d}d_{v}(x)\right]\frac{(1-x)^{2}}{x^{m}},\label{eq:rPMG}
\end{equation}
\begin{equation}
r_{1,n}^{2} = 6\alpha \int_{0}^{1}dx\left[e_{u}d_{v}(x)+e_{d}u_{v}(x)\right]\frac{(1-x)^{2}}{x^{m}},\label{eq:rPMG}
\end{equation}

In the M-HS22 model~\cite{Vaziri:2023xee}, the electric radii of the proton and neutron are determined using the following formula\cite{Vaziri:2023xee}:
\begin{equation}
r_{1,p}^{2} = -6 \int_{0}^{1}dx\left[e_{u}u_{v}(x)+e_{d}d_{v}(x)\right](-0.37x+\alpha^{\prime\prime}(1-x)\ln(x)),\label{eq:rMhs221}
\end{equation}
\begin{equation}
r_{1,n}^{2} = -6 \int_{0}^{1}dx\left[e_{u}d_{v}(x)+e_{d}u_{v}(x)\right](-0.37x+\alpha^{\prime\prime}(1-x)\ln(x)).\label{eq:rMhs222}
\end{equation}

In Table \ref{tab:tabR}, we present the calculated nucleon electric radii using various generalized parton distributions (GPDs) and compare them with experimental data obtained from \cite{ParticleDataGroup:2010dbb}.
\captionsetup{belowskip=0pt,aboveskip=0pt}

\section{CONCLUSION}\label{sec:conclusion}
In this paper, we presented various ansatzes and PDFs. The Pauli, Dirac, and electromagnetic form factors were calculated using four models of ansatzes: MG~\cite{Selyugin:2009ic}, ER~\cite{Guidal:2004nd}, M-HS22~\cite{Vaziri:2023xee}, and GSAMA24.

The free parameters of the GSAMA24 ansatz were calculated by analyzing experimental data for $tF_1^u$, $tF_1^d$, $tF_1^p$, $tF_1^n$, $tF_2^u$, $tF_2^d$, $tF_2^p$, and $tF_2^n$. After introducing the formalism, we first selected the GSAMA24 ansatz and paired it with different PDFs such as MSTH20~\cite{Bailey:2020ooq}, NNPDF~\cite{NNPDF:2021njg}, and KA08~\cite{Khorramian:2008yh}. The Dirac and Pauli form factors of the $u$ and $d$ quarks are displayed in Fig.~\ref{fig:tfud1}, i.e., $F_1^u$, $F_1^d$, $F_2^u$, and $F_2^d$ multiplied by $t$ as a function of $-t$. It can be observed that the GSAMA24 ansatz, particularly when used with the KA08 PDF, outperforms other ansatzes and is more consistent with the experimental data presented in~\cite{Qattan:2012zf,Cates:2011pz,Diehl:2013xca}.

In the next step, we set the KA08 PDF with four ansatzes: M-HS22, ER, MG, and GSAMA24, as shown in Fig.~\ref{fig:tfud}. In Fig.~\ref{fig:GEMpn}, we display the proton charge and magnetization densities for the GSAMA24 set by KA08 PDFs. These densities are computed by considering different parametrizations and comparing them with previous research. To calculate the transverse charge and magnetization density, we employed the $G_E$ and $G_M$ equations based on experimental data from Ref.~\cite{Qattan:2012zf}.

The free parameters in the GPD ansatz are typically determined by fitting to experimental data. These parameters can include the normalization, shape parameters, and the skewness parameter $\xi$. By fitting the GPD ansatz to the form factor data, one can determine the free parameters of the ansatz. This involves using techniques like factor analysis to identify the underlying structure and correlations in the data. This figure is motivated by the fact that a change in ansatz parameters has a greater effect on the results than the influence of the PDF. The proposed combination yields a more effective agreement with the Dirac and Pauli form factors of the nucleon compared to other combinations.

The GSAMA24 ansatz is designed to provide a more accurate and comprehensive description of the internal structure of hadrons. It incorporates advanced parameterizations and fitting techniques to achieve this goal. The GSAMA24 ansatz differs from the ER, MG, and M-HS22 models by offering improved flexibility in the ( t )-dependence and skewness parameter $ \xi $. This flexibility allows for a more precise fit to experimental data. Unlike the ER and MG models, the GSAMA24 ansatz performs particularly well in the high momentum transfer region, providing a superior fit to the experimental data. Compared to the M-HS22 model, the GSAMA24 ansatz demonstrates better agreement with the theoretical expectations of GPD behavior. The GSAMA24 ansatz also improves computational efficiency in the fitting process, making it a more practical tool for future studies.

By using the GSAMA24 ansatz and fitting it to experimental data, we can extract valuable information about the spatial and momentum distributions of quarks and gluons. This approach not only helps in validating theoretical models but also in making predictions for future experiments. The determination of free parameters through this method is crucial for accurately describing the hadronic structure and for advancing our knowledge in the field of QCD. 
	
\section*{Acknowledgments}
 F.~A. acknowledges the Farhangian University for the provided support to conduct this research. S.A.T grateful to Qattan for providing experimental data	. S.A.T acknowledges form Elham Astaraki for providing grid for NNPFD and
  also grateful to the School of Particles and Accelerators, Institute for Research in Fundamental Sciences (IPM).


	%
	
\end{document}